# Reflective Liquid-Crystal Phase Shifter based on Periodically Loaded Differential Microstrip Lines

Yuh-Chyi Chang , Tien-Lun Ting , Pei-Ru Chen , and Tsung-Hsien Lin

*Abstract*— Beamforming technologies have garnered substantial attention for their capacity to manipulate the wireless propagation environment, establishing them as a pivotal technology for enhancing capacity and coverage in communication networks. High-performance phase control unit is a key element for beamforming. This paper details the design and fabrication of a 3.5GHz reflective liquid-crystal (LC) phase shifter. This component is formulated from a differential pair of transmission lines, periodically loaded with floating electrodes, wherein the LCs in the overlapping area serve as a variable capacitor. By applying AC to the permittivity in the overlapping area, the degree of phase shift can be altered. Both simulation and measurement results display impressive Figures of Merit (FoM) of 101.3°/dB and 85.7°/dB, respectively. The grounding issue inherent to CPWs on glass substrates is adeptly mitigated by utilizing the virtual ground present within a differential pair configuration. This strategy not only enables the low-cost manufacturing of the phase shifter array but also significantly facilitates the development of beamforming in practical applications.

## I. INTRODUCTION

The escalating demand for higher data rates has catalyzed the swift development of wireless communication systems [1]-[2]. Consequently, carrier frequencies have been elevated to secure wider bandwidths for massive data transmission. However, these higher frequencies introduce challenges such as significant propagation loss and high directivity. To navigate these challenges, Beamforming technologies, such as Phased Array Antenna (PAA), and Reconfigurable Intelligent Surfaces (RISs) have been implemented, garnering substantial attention in recent years [3]-[4]. PAA and RIS is crucial for optimizing beamforming at high frequencies like mmWave, Radar, and 5G-6G. By adjusting the phase difference between each radiating and receiving units, it modifies the direction and shape of signals without physical movement. Therefore, beamformaing is considered a powerful technologies in future communication systems, improving area coverage, and enhancing signal patterns for better adaptability to varying system requirements. Phase shifter is a pivotal component in the beamforming implementation, which can be achieved in several ways, including RF MEMS [5]-[7], semiconductor solutions [8]-[10], ferroelectrics [11] (e.g., barium-strontium-titanate, BST), and liquid crystal technologies.

Liquid crystal (LC) materials offer numerous advantages in the fabrication of phase shifters [12]-[16]. One notable benefit is their minimal power consumption required to alter or maintain the LC state. For instance, a mere 0.1W is necessary to drive the LC in a 32-inch panel. Additionally, LCs exhibit low dielectric loss. Merck, a major supplier of LC materials, has pioneered LicriOn™ technology, which prioritizes LC materials suitable for microwave applications. This innovation diminishes the tangential loss of LC to less than 0.01, expanding its applicability across a wide array of applications. Concerning manufacturing costs, the liquid crystal display (LCD) process, which has been mature and widely utilized for decades, presents a viable and cost-effective method for producing microwave phase shifters. Moreover, we can realize high-transparency RIS with metal mesh patterns on glass generated through the lithography process, given the advantage to seamlessly integrated into urban landscapes for commercial purposes. In essence, the LC-based phase shifter emerges as a promising solution.

Several types of LC-based phase shifter have been proposed previously [17]. These encompass two principal structures for loaded transmission lines: the microstrip line (MSL) with bulk-like liquid crystal [18]-[23], and the coplanar waveguide (CPW) periodically loaded by shunt liquid-crystal varactors [24]. For MSL, the shunt capacitance of the microwave is dictated by the liquid crystal orientation, thus its tunability is constrained by the dielectric properties of the liquid crystal material. The considerable thickness of MSL also yields a slow liquid crystal response, rendering it unsuitable for phased arrays, which necessitate a rapid operational response. In contrast, CPW presents a better solution due to its reduced cell gap to several micrometers, restoration of response time to milliseconds. However, ensuring the CPW's GND electrode is effectively grounded proves challenging when utilizing glass substrates, as LCD manufacturing does not offer a mature Through Glass Vias (TGV) hole process. Hence, an alternative solution to the periodically loaded CPW is requisite.

In this study, we formulated a novel LC phase shifter design based on a differential pair of microstrip lines periodically loaded with floating electrodes, where the overlapping area

The work was supported in part by the National Science and Technology Council of Taiwan under grant NSTC 112-2218-E-110-004, NSTC 112-2223-110-004, and by the Sixth Generation Communication and Sensing Research Center funded by the Higher Education SPROUT Project, Ministry of Education of Taiwan. *(Corresponding author: Tsung-Hsien Lin).*

Yuh-Chyi Chang (e-mail: changtrisha23@gmail.com), Tien-Lun Ting (e-mail: tienlun@gmail.com), Pei-Ru Chen (e-mail: peiru8704@gmail.com), and Tsung-Hsien Lin (e-mail: jameslin@mail.nsysu.edu.tw) are with the Department of Photonics, National Sun Yat-sen University, Kaohsiung 80424, Taiwan.

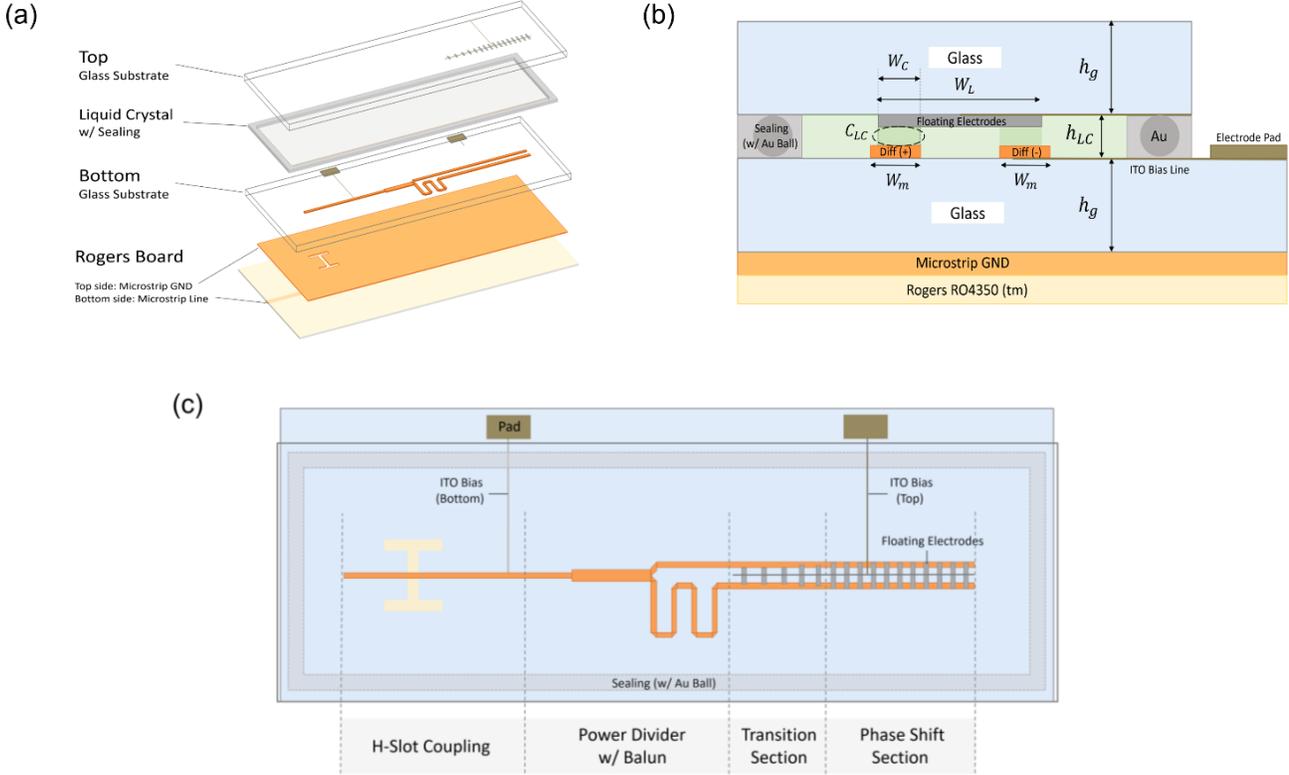

**Fig. 1.** Configuration of the proposed LC Phase Shifter (a) 3D Layer View (b) Cross-side view of the phase shift section, Dimensions: Microstrip line width, $W_m$ = 0.6 mm ; Floating electrode dimensions, $W_L$ = 2.9 mm, $l_L$ = 0.5 mm, $h_L$ = 3 μm; Overlapping area width, $W_c$ = 0.6 mm ; Liquid crystal thickness, $h_{LC}$ = 12 μm ; Glass thickness, $h_g$ = 0.5 mm.

controls the LCs, serving as a variable capacitor. Phase shifting is achieved by applying different AC voltages to alter LC permittivities. With the virtual ground existing between differential pair, grounding floating electrodes is no longer needed, resulting in simplified process compared with the typical CPW. As for liquid crystal materials, there exists a dipolar resonance near 1GHz. It implies high tangent loss around that band. To ensure the feasibility of liquid crystal based phase adjustment, we design a phase shifter operating at 3.5 GHz to evaluate the performance of liquid crystals in sub-6GHz of FR1. This is to facilitate the possibility of future adaptation for higher band of FR1 such as 6.425 to 7.125GHz, which has been assigned as the next-generation mobile communication band. Simultaneously, developing a scalable structural design for mass production.

## II. DESIGN AND ANALYSIS OF THE LC PHASE SHIFTER

Figure 1 presents the structure of the proposed liquid crystal phase shifter, a reflective type, one-port component. The phase shift section consists of a differential pair of microstrip lines periodically loaded with shunt LC variable capacitors, incorporating a total of 11 units of floating electrodes. To create a differential pair from a 50 Ω microstrip line for RF signal input, a power divider with Balun is implemented, wherein the transition section facilitates Chebyshev impedance matching between loaded and unloaded differential pairs. Given the fragility of glass and to circumvent impedance discontinuities introduced by soldering SMA, a copper microstrip line for RF signal input is plated on the bottom surface of the Rogers RO4350 substrate. Concurrently, a copper microstrip ground is positioned on the top surface, with an H-slot designed to couple the signal into the microstrip line situated between glass substrates. Figure 1(b) depicts the cross-side view of a unit in the phase shift section. It is implemented using two glass substrates, each with a thickness of 0.5mm, and infused with nematic liquid crystals. The two 50 Ω microstrip lines with a width of 0.6mm are placed on the bottom glass substrate, forming a differential pair, while the floating electrodes, with dimensions of ($W_L$, $l_L$, $h_L$) in the (X, Y, Z) direction, are situated on the top glass substrate. All metals are made of copper with a thickness of 3 μm. Voltage is applied through the electrode pad on the bottom substrate and transmitted to microstrip lines via an ITO bias line with a width of 10 μm. To deliver bias to the upper floating electrode, Au balls are mixed into the epoxy sealing to contact the ITO bias line on the top glass substrate with the electrode pad on the bottom, simultaneously providing a uniform cell gap with a thickness of 12 μm. The overlapping area between the floating element and the differential pair serves as the LC variable capacitor, designated with a capacitance value of $C_{LC}$.

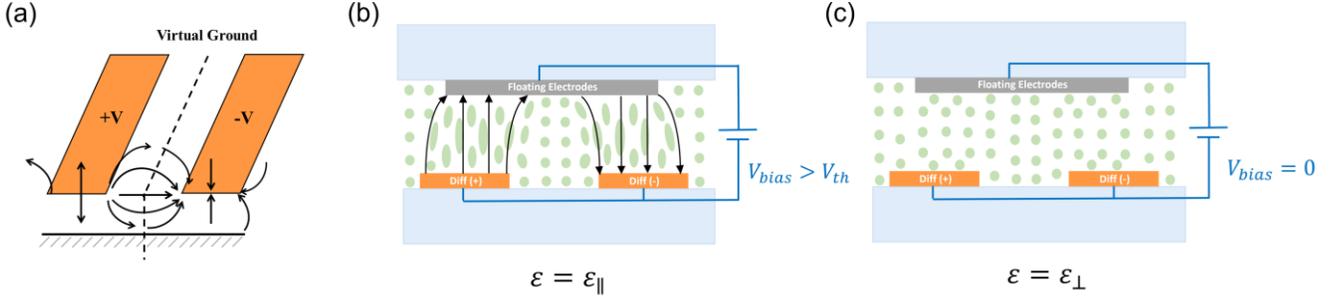

**Fig. 2.** (a) Electric field distribution of a differential pair. Schematic diagram of LC molecules (b) with bias voltage (c) without bias voltage applied

TABLE I
MATERIAL PROPERTIES

| Material | | Permittivity | Loss | Conductivity |
|---|---|---|---|---|
| LC | $\varepsilon_\perp$ | 2.5 | 0.007 | NA |
| | $\varepsilon_\parallel$ | 3.6 | 0.003 | NA |
| Copper | | 1 | 0 | 58000000 |
| Glass | | 5.4 | 0 | NA |

*A. Phase Shift Section design*

In a transmission line model, LCs can function as a variable capacitor, wherein permittivity variation modulates capacitance. This modulation of the RF signal's phase velocity enables phase shift adjustment by merely altering the external bias magnitude, realizing an LC phase shifter. Additionally, the minimal power required to modify or sustain the LC state establishes it as an energy-efficient solution for passive phase shifter design.

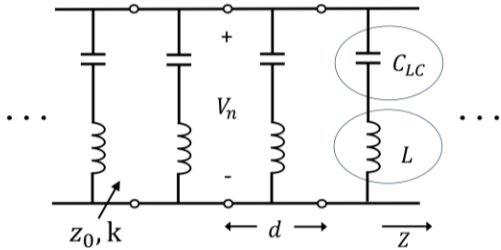

**Fig. 3.** Equivalent circuit of the phase shift section.

Referencing the previously discussed CPW structure, we formulated a new LC phase shifter design based on a differential pair of microstrip lines periodically loaded by shunt LC varactors. Figure 2(a) displays the electric field distribution in a differential pair, transmitting signals via two microstrip lines of equal amplitude but opposing phases, operating in odd mode. Owing to the presence of a Virtual Ground at its center, grounding the floating electrode is unnecessary, an aspect beneficial to the LCD manufacturing process. Figure 2(b) shows the orientation of the liquid crystals when a voltage is applied, with the liquid crystals parallel to the electric field generated in the overlapping region of the floating electrode (gray) and signal electrodes (orange). Given that the quasi-TEM wave propagates in the transmission line parallel to the glass substrate, the effective permittivity perceived is $\varepsilon_\parallel$. In the absence of bias, presented in Figure 2(c), the orientation of liquid crystals is dictated by the homogeneous alignment layer, therefore the effective permittivity perceived is $\varepsilon_\perp$. As LCs are tuned, the size of overlapping area becomes a crucial design factor, influencing capacitance value, C, and phase velocity, $\beta$, given by the following formula:

$$V_p = \frac{\omega}{\beta} = \frac{1}{\sqrt{LC}} \quad (1)$$

To facilitate the analysis and design of the proposed phase shifter, a circuit model was developed to derive theory and equations, serving as design reference points. Figure 3 illustrates the equivalent circuit model of the phase shift section, representing a transmission line periodically loaded with shunt varactors. The transmission line possesses a characteristic impedance of $Z_0$ and a propagation constant of k. The periodically shunt impedance Z is composed of a liquid crystal capacitance $C_{LC}$ and an inductance $L$ introduced by a small segment of floating electrodes, which is given by:

$$Z = \frac{1}{j\omega C_{LC}} + j\omega L = \frac{1 - \omega^2 L C_{LC}}{j\omega C_{LC}} \quad (2)$$

Ensuring the propagation of RF signals requires the shunt branch to function as a capacitive impedance, demanding that the numerator of the equation, $1 - \omega^2 LC$, be both greater than zero and less than one. The frequency at the turning point of capacitive and inductive impedance is defined as the cutoff frequency $f_C$. Therefore, the design of component must accommodate the operating frequency as the electromagnetic wave cannot propagate when the branch operates as an inductive impedance. The periodically loaded lines, depending on the frequency and normalized susceptance values, may exhibit either a passband or stopband and thus can be regarded as a type of filter. The Bloch impedance $Z_B$ and Bloch frequency $f_B$, demarcate the characteristic impedance and passband of the periodic structure. Consequently, we must ensure that the operating frequency does not closely approach the Bloch frequency, beyond which the periodic structure is unable to transmit signals.

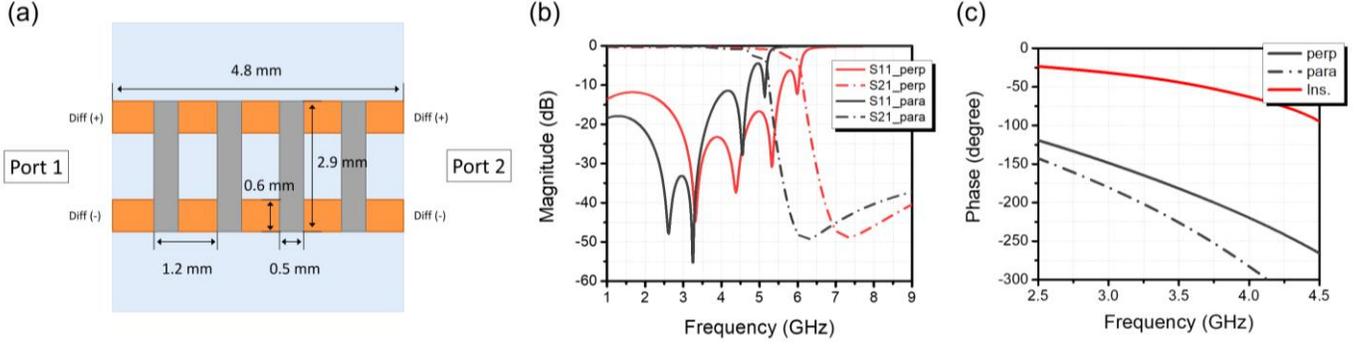

**Fig. 4.** (a) Schematic of a 4-unit phase shift section (b) Simulated results for return loss (S11) and insertion loss (S21) (c) phase shift

Subsequently, a phase shifter operating at 3.5GHz was designed, with simulations executed using HFSS and material properties presented in Table 1. Initially, to account for the fringe field, we simulated a pitch of the proposed structure in Ansys Q3D to obtain the resultant capacitance ($C_{eff}$) and inductance ($L$). Then, we calculated the corresponding Bloch frequencies ($f_B$) and Bloch impedances ($Z_B$) under different liquid crystal states using ABCD matrix [25], as detailed in Table 2. Considering that LC modulation might induce a change in capacitance value, which could, in turn, cause a frequency drop, we designed the cutoff frequency ($f_C$) to be twice the operating frequency.

TABLE II
DESIGN PROPERTIES FOR PHASE SHIFT SECTION

| LC Permittivity ($\varepsilon_{eff}$) | Effective Capacitance ($C_{eff}$) | Bloch Impedance ($Z_B$) | Bloch Frequency ($f_B$) | Cutoff Frequency ($f_C$) |
|---|---|---|---|---|
| 2.5 ($\varepsilon_\perp$) | 1.05 pF | 17.5 Ω | 15 GHz | 7.3 GHz |
| 3.6 ($\varepsilon_\parallel$) | 1.46 pF | 15 Ω | 13 GHz | 6.2 GHz |

Based on the design, we first simulated a phase shift section comprising 4 units, totaling 4.8 mm. The structural dimensions are presented in Figure 4(a), wherein the distance between the differential pair and the overlapping area of the floating electrode primarily influences the inductance and capacitance values. The resultant characteristic impedance for the phase shift section is approximately 28 Ω; contrastingly, the characteristic impedance for unloaded differential pair is 91 Ω, signaling the impedance drop induced by loading capacitors.

Figure 4(b) illustrates the simulated return loss (S11) and insertion loss (S21) under different LC states, where 'perp' (perpendicular state, $\varepsilon_\perp$) denoting results without voltage applied and 'para' (parallel state, $\varepsilon_\parallel$) with voltage applied. At the center frequency of 3.5 GHz for both permittivities, S21 hovers around -0.4 dB with S21 below -15 dB. The insertion loss experiences a rapid surge from 5 GHz to 6 GHz, attributed to the transition from capacitive to inductive impedance, which implies the location of the cutoff frequency. Upon the application of voltage, the liquid crystal permittivity shifts from a perpendicular to a parallel state, resulting in the cutoff frequency transitioning from 7.3 GHz to 6.2 GHz. Figure 4(c) presents the accumulated phase under LC 'perp' and 'para' state and the phase difference obtained by their subtraction, indicating the phase shift degree of the structure. The accumulated phase difference is approximately 44 degrees for a 4-unit periodic structure, and the total size is 4.8 mm. Consequently, by adjusting the number of loaded pitches, the desired phase shift degree can be achieved. The component sizes, being much smaller than the dielectric wavelength, can be regarded as a lumped circuit, thereby mitigating concerns regarding the capacitive effect at higher frequencies.

*B. Transition Section design*

As verified by Eq.1 and simulation results for the phase shift section, loading capacitance results in a decrease in the characteristic impedance, so designing a transition section matching from 91 Ω to 28 Ω is necessary to avoid strong reflections caused by impedance mismatch. Various matching methods have their advantages and disadvantages. Considering the narrow bandwidth of quarter-wavelength might limit the frequency range of the phase-shifted section, we opted for the Chebyshev impedance transformation method. Originating from the theory of small reflections depicted in Figure 5, the Chebyshev impedance transformer establishes a broader bandwidth matching network. A larger number of sections provides superior matching effects but also demands a correspondingly longer distance to establish the matching network. Therefore, optimal matching within restricted spatial constraints requires judicious trade-offs.

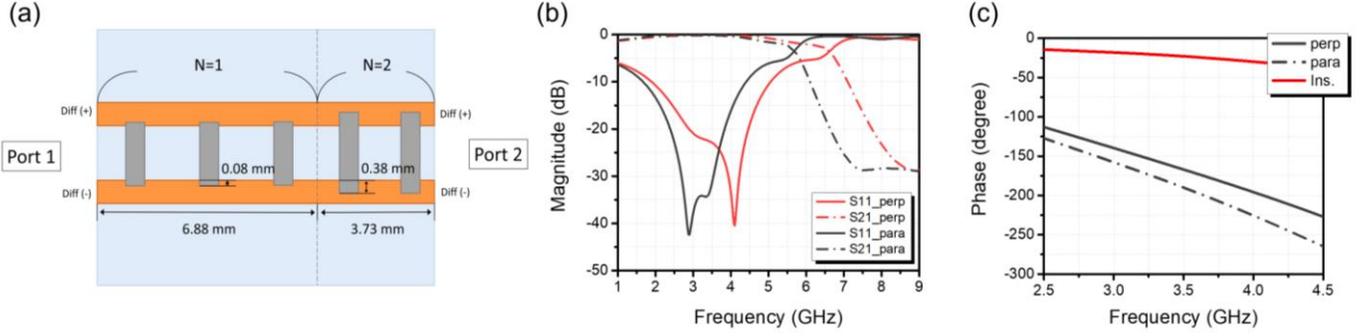

**Fig. 6.** (a) Schematic of the transition section (b) Simulated results for return loss (S11) and insertion loss (S21) (c) phase shift

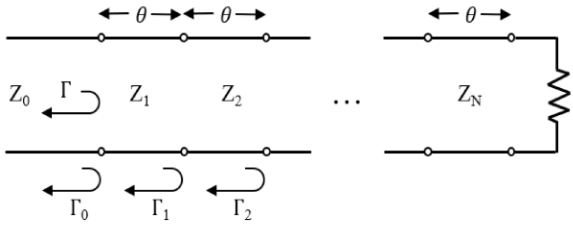

**Fig. 5.** Theory of small reflections.

In this work, a two-section ($N = 2$) impedance matching is employed [25]. We determine the maximum allowable reflection coefficient magnitude in the passband ($\Gamma_m = 0.05$) and obtain the ratio of target impedances ($Z_L / Z_0$) to derive the reflection constant ($\Gamma_n$) and impedance of each section ($Z_n$). As follows:

$$\Gamma(\theta) = 2e^{-j2\theta}\left[\Gamma_0 \cos\theta + \frac{b}{2}\cos\theta - \frac{b}{2}\right] \quad (3)$$

$$= Ae^{-j2\theta}T_2(\sec\theta_m \cos\theta)$$

where is considered,

$$\sec\theta_m = \cosh\left[\frac{1}{N}\cosh^{-1}\left(\frac{\ln Z_L / Z_0}{2\Gamma_m}\right)\right] \quad (4)$$

Once is known, the fractional bandwidth can be calculated as,

$$\frac{\Delta f}{f_0} = 2 - \frac{4\theta_m}{\pi} \quad (5)$$

The characteristic impedances can be found by using the approximation,

$$\Gamma_n \cong \frac{1}{2}\ln\frac{Z_{n+1}}{Z_n} \quad (6)$$

After determining the required impedance for each section, simulations were conducted in Ansys Q3D by varying the overlapping size of the structure, enabling us to acquire the resultant capacitance and inductance values that closely aligned with the target impedances. Table 3 details the calculated impedance ($Z$) and effective capacitance ($C_{eff}$) for each section. Subsequently, we replaced the port impedance with the desired values and optimized the structure based on the performance criteria of return loss and insertion loss. Figure 6(a) provides a schematic of the structure, showcasing the size of the overlapping area and the distances between each section.

TABLE III
DESIGN PROPERTIES FOR TRANSITION SECTION

| SECTION | Impedance (Z) | Effective Capacitance ($C_{eff}$) | Number of pitch | $C_{eff}$ per pitch |
|---|---|---|---|---|
| N=1 | 31.94 Ω | 1.13 pF | 3 | 0.37 pF |
| N=2 | 17.31 Ω | 3.53 pF | 2 | 1.76 pF |

In Figure 6(b), the simulated return loss (S11) and insertion loss (S21) of the proposed structure are displayed under two states, with "perp" and "para" representing results without and with voltage applied, respectively. At 3.5 GHz, S11 is approximately -0.2 dB, while S21 is below -20 dB. The simulated bandwidth for -25 dB is roughly 1.2 GHz, a slight deviation from the calculated result using Eq. 5, which is about 1.74 GHz. Since the Chebyshev Impedance Matching section is also implemented with a periodically loaded structure, it adapts to the varying impedance of the Phase Shift section when voltage is applied, thereby providing phase-shifting effects throughout the structure. As depicted in Figure 6(c), this results in an overall phase shift of 22 degrees. To verify the importance of transition section, we simulated the proposed phase shifter with and without transition section. The absence of transition section results in significant reflection at the interface due to impedance mismatch, leading to the emergence of multiple pronounced resonance dips. Hence, the inclusion of an impedance matching section is necessary.

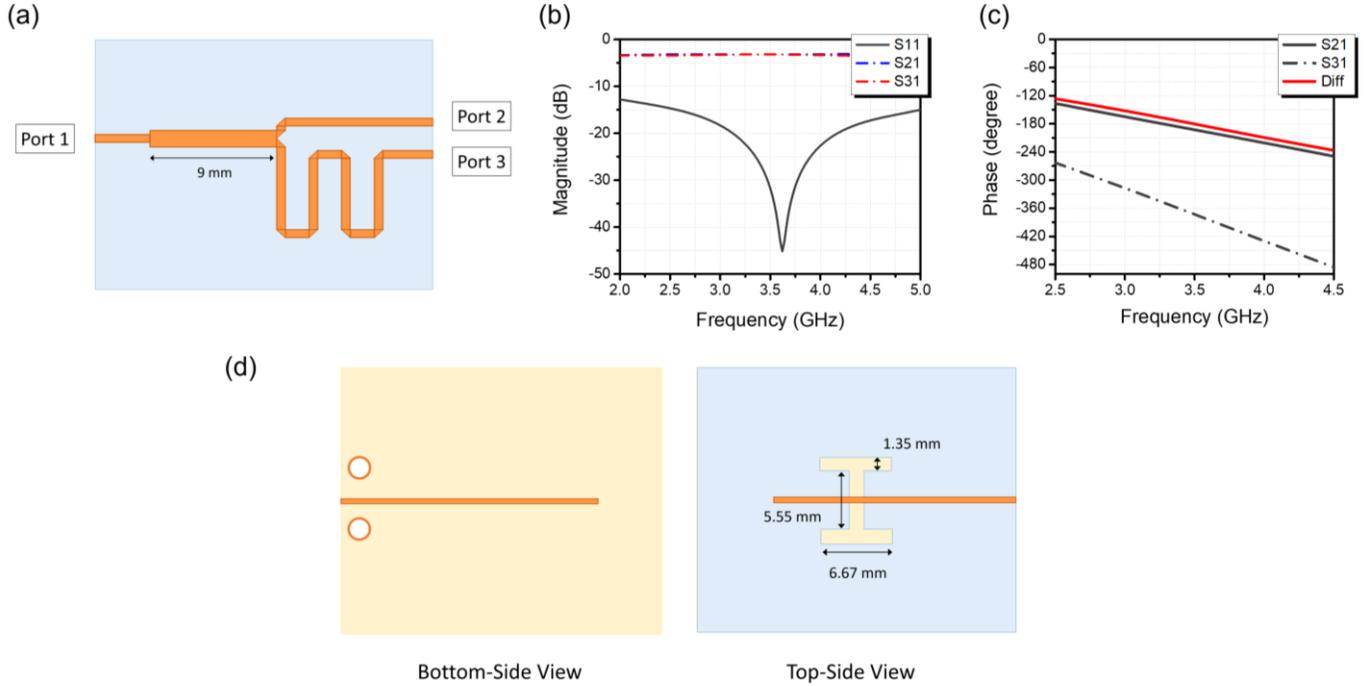

**Fig. 7.** (a) Schematic of the Balun section (b) Simulated results for return loss (S11) and insertion loss (S21) (c) phase difference (d) Schematic of H-slot Coupling Section

*D. Power Divider Balun & H-slot Coupling Design*

To create a differential pair from a 50 Ω microstrip line for RF signal input, a power divider with Balun was implemented. Figure 7(a) illustrates how the power divider bifurcates a single 50 Ω microstrip line into two, incorporating a quarter-wavelength impedance matching to transition from 50 Ω to 100 Ω in the interim. This configuration attains a 180-degree phase difference at 3.5GHz via disparate path lengths of the two lines, thereby effectively forming a differential pair. The simulated results are depicted in Figure 7(b) and (c). A comparison of the transmission coefficient at port 2 (S21) and port 3 (S31) reveals that the power is almost equally divided, with a mere 0.03 dB discrepancy. This minor difference is deemed acceptable, accounting for losses due to varying path lengths and impedance mismatches at corners.

Since the LC phase shifter is deployed on glass substrates, an H-slot coupling structure was designed to enable signal coupling from a 50 Ω microstrip line on the bottom side of the Rogers substrate attached with SMA connector to a 50 Ω microstrip line sandwiched in the glass substrate. The dimensions of this structure are provided in Figure 7(d).

### III. EXPERIMENTAL RESULT AND DISCUSSION

To validate the feasibility of the proposed LC phase shifter structure, we fabricated a prototype to assess its performance. The pertinent structure dimensions and material properties are described previously in Section 2. The LC Phase Shifter, as demonstrated in Figure 8(a), is assembled by the Rogers substrate and the glass cell. The Rogers substrate is copper-plated on both surfaces: the top features a Microstrip GND with an H-slot pattern, while the bottom comprises a Microstrip line. The glass cell is fabricated through an LCD manufacturing process. The glass cell is fabricated using LCD manufacturing process. First, a thin layer of ITO (Indium Tin Oxide) is deposited on glass via Physical Vapor Deposition (PVD), followed by photolithography for patterning through light etching, and annealing to create the ITO electrode pattern. Copper is then plated by sputtering and wet etching methods. Alignment layer is applied and rubbed to create grooves that define the alignment direction of liquid crystal molecules. Sealing material is blended with Au balls and applied along the edges to achieve a uniform cell gap and establish conductivity between the upper and lower ITO electrodes. Liquid crystals (GT7-29001, Merck) were spread onto the substrate and subjected to vacuum lamination to assemble the glass substrates, forming the glass cell. Upon stacking both components, the Microstrip line connects to a 50 Ω SMA connector, serving as the RF signal input interface. For the measurement setup, an RF Vector Network Analyzer (VNA) from Keysight (model P9384B) is utilized to measure the S-parameter and phase characteristics. An AC bias voltage (1kHz, square wave) was applied through Function Generator (Agilent, model 33220A), connecting to the electrode pads via a crocodile clip.

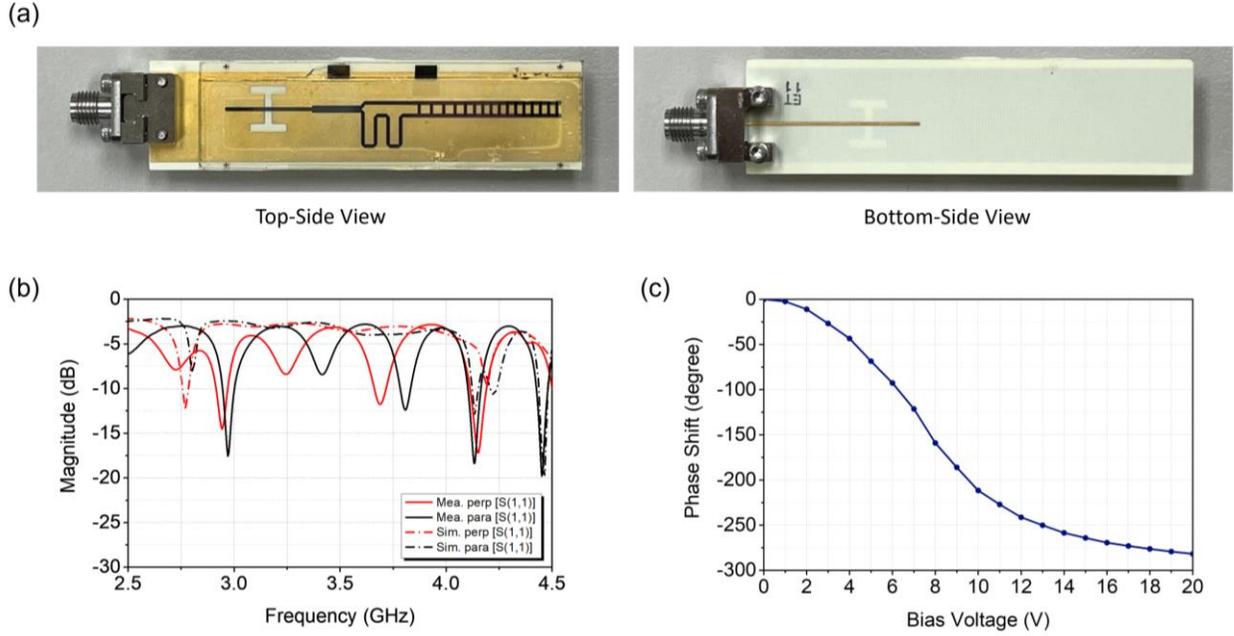

**Fig. 8.** (a) Figuration of the proposed LC Phase Shifter (b) Simulated and Measured Return Loss (S11) of the proposed LC Phase Shifter (c) Measured Phase-Shift Degree vs. Bias Voltage at 3.5 GHz

Figure 8(b) displays resultant reflection coefficient (S11) for the proposed reflective-type LC phase shifter, where solid lines indicate measurement results and dashed lines represent simulation results. The red lines, labeled "perp," correspond to the device without voltage applied, while the black lines, labeled "para," represent the device with voltage applied. TheLC Phase shifter features reflection coefficient above -10 dB from 2.8GHz to 4.2GHz, where the device's open-ended configuration introduces dips attributed to strong reflections from resonances within the structure. However, the additional dips are not anticipated to significantly impact future application scenarios, especially in massive Reconfigurable Intelligent Surface (RIS) applications. In high element number RIS applications, the impact of high-loss intervals can be effectively mitigated through 1-bit operations.

Figure 8(c) depicts the relationship between the measured phase shift degree and the applied bias voltage, spanning from 0V to 20V. The phase shift commences at 2V, denoting the LC material's threshold voltage. A notable and abrupt change in the phase shift degree occurs between 4V and 12V. With further voltage increase, the curve gradually plateaus, showcasing the characteristic behavior of LC material under an electric field. To evaluate the performance of a phase shifter, the classical figure of merit (FoM) [26] is employed, defined as the ratio between the maximum differential phase shift and the maximum insertion loss, expressed as:

$$FoM = \frac{\Delta \Phi_{max}}{IL_{max}} \qquad (7)$$

The maximum measured phase shift degree is 282 degrees, with a reflection of -3.29 dB. This configuration results in a FoM of 85.7°/dB for the entire component, measuring 84.5mm.

In simulations, the maximum phase shift degree attains 320 degrees with a reflection of -3.13 dB, yielding a higher FoM of 102.2°/dB. The differences between measured and simulated results are due to several reasons. Possible causes include fabrication process-induced structural errors, causing the cell gap to be larger than expected, contributing to the capacitance difference, which can be observed in the slight discrepancy of dips. As well as imperfections in the alignment of the liquid crystal materials and uncertainties in the dielectric constants. Further research will be conducted in the future to address these details to optimize this liquid crystal phase shifter.

We have successfully demonstrated the viability of implementing this structure within the sub-6GHz FR-1 framework, mitigating the high loss and resonance issues of liquid crystals at lower frequencies. Building upon this research findings, our future endeavors will involve the realization of Reflective Intelligent Surface (RIS) applications in higher band of FR-1 or even FR-3. Recognizing that coupled and balun segments contribute to an overall component size that may surpass the physical limitations of the phase array unit, we plan to harness the inherent advantage of dipole antennas, specifically their 180-degree phase difference. This strategic utilization will be incorporated into the existing architecture of the phase-shifted segment to address size constraints and enhance overall system performance.

IV. CONCLUSION

In this paper, the development of a reflective 3.5GHz LC phase shifter has been presented. The phase shifter, constructed using a differential pair transmission line periodically loaded with floating electrodes, employs the overlapping area as a variable capacitor to modulate the permittivity of the liquid crystals. This innovative approach adeptly mitigates the grounding issue commonly encountered with coplanar waveguides (CPW) by utilizing the inherent virtual ground between the differential pair. A thorough overview of the theory and design of each section, encompassing the Phase-Shift Section, Chebyshev Impedance Matching, Power Divider with Balun, and H-slot Coupling, has been provided. Simulations revealed a maximum phase shift of 320 degrees for the entire LC phase shifter, while experimental measurements exhibited a phase shift of 282 degrees. Consequently, the Figures of Merit (FoM) are 101.3°/dB and 85.7°/dB for the simulated and experimental results, respectively, with the overall component size measuring 84.5 mm. The variance between the measured and simulated phase shifts can be attributed to our limited fabrication capability and the precise control of the liquid crystal material. It is pivotal to underscore that our phase-shifter design, being compatible with panel factory processes, possesses the potential to markedly reduce manufacturing costs. Furthermore, it simplifies the implementation of beamforming, enhancing their accessibility and practicality for future applications. We successfully validated the feasibility of incorporating this structure in the sub-6GHz FR-1 framework, addressing high loss and resonance challenges in liquid crystals at lower frequencies. Future efforts will focus on implementing Reflective Intelligent Surface (RIS) applications in higher band of FR-1or FR-3 based on these research findings.